\documentclass[prl,twocolumn,showpacs,preprintnumbers,amsmath,amssymb]{revtex4-1}

\usepackage{graphicx}
\usepackage{dcolumn}
\usepackage{bm}
\usepackage{comment}
\usepackage{bbold}

\begin{document}

\title{Anomalous edge state in a non-Hermitian lattice}
\date{\today}

\author{Tony E. Lee}
\affiliation{Department of Physics, Indiana University Purdue University Indianapolis (IUPUI), Indianapolis, Indiana 46202, USA}

\begin{abstract}
We show that the bulk-boundary correspondence for topological insulators can be modified in the presence of non-Hermiticity. We consider a one-dimensional tight-binding model with gain and loss as well as long-range hopping. The system is described by a non-Hermitian Hamiltonian that encircles an exceptional point in momentum space. The winding number has a fractional value of 1/2. There is only one dynamically stable zero-energy edge state due to the defectiveness of the Hamiltonian. This edge state is robust to disorder due to protection by a chiral symmetry. We also discuss experimental realization with arrays of coupled resonator optical waveguides.
\end{abstract}

\maketitle

\emph{Introduction.---}
The bulk-boundary correspondence is a central principle that governs the band structure of tight-binding models \cite{ryu02,asboth16}. It says that the bulk of a lattice is characterized by a topological invariant, whose value determines the existence of gapless states that are localized on the edges of the sample. The bulk-boundary correspondence applies universally to all noninteracting tight-binding models, which are usually assumed to be Hermitian, i.e., closed systems. As such, it has been used to predict edge states in a variety of settings, including solid state \cite{hasan10,qi11}, cold atoms \cite{aidelsburger13,miyake13}, photonics \cite{kraus12,rechtsman13,zeuner15}, and even acoustics \cite{yang15,susstrunk15}.

In a one-dimensional tight-binding model, the topological invariant is the winding number, which is always an integer \cite{ryu02,asboth16}. If the winding number is 0, there are no edge states. If the winding number is $\pm1$, there will be a pair of zero-energy edge states (one on the left and one on the right). As the Hamiltonian is modified, the edge states are ``topologically protected'' because the winding number changes only when the gap closes.




The bulk-boundary correspondence was developed for Hermitian systems, as motivated by solid state experiments. However, experiments with photonics are intrinsically non-Hermitian due to gain and loss, which raises the question of whether they can display physics beyond the bulk-boundary correspondence. The general conclusion so far is that the usual bulk-boundary correspondence still holds in the presence of non-Hermiticity, although the spectrum may be complex \cite{rudner09,hu11,esaki11,diehl11,zhu14,malzard15,joglekar15,sanjose16}.

In this Letter, we show that the bulk-boundary correspondence is modified in the presence of non-Hermiticity. We consider a one-dimensional tight-binding model with gain and loss, motivated by recent experiments with optical waveguides \cite{guo09,ruter10,kraus12,rechtsman13,regensburger12,zeuner15,zhen15}. First, we show that the winding number can have a fractional value of 1/2, because the Brillouin zone is $4\pi$ periodic when circling an exceptional point (a non-Hermitian degeneracy) \cite{heiss12}. An open chain exhibits a zero-energy eigenvalue, which is peculiar because it is defective \cite{golub96}. Although there are two edge states, only one of them is dynamically stable. This edge state is protected by a chiral symmetry and is robust to disorder until the band gap closes. These new features are due to the fact that a non-Hermitian matrix can be defective, while a Hermitian matrix is always diagonalizable \cite{golub96}. We also discuss experimental implementation with optical resonators and waveguides.


\begin{figure}[t]
\centering
\includegraphics[width=3.3 in,trim=3.4in 1.9in 1.5in 1.9in,clip]{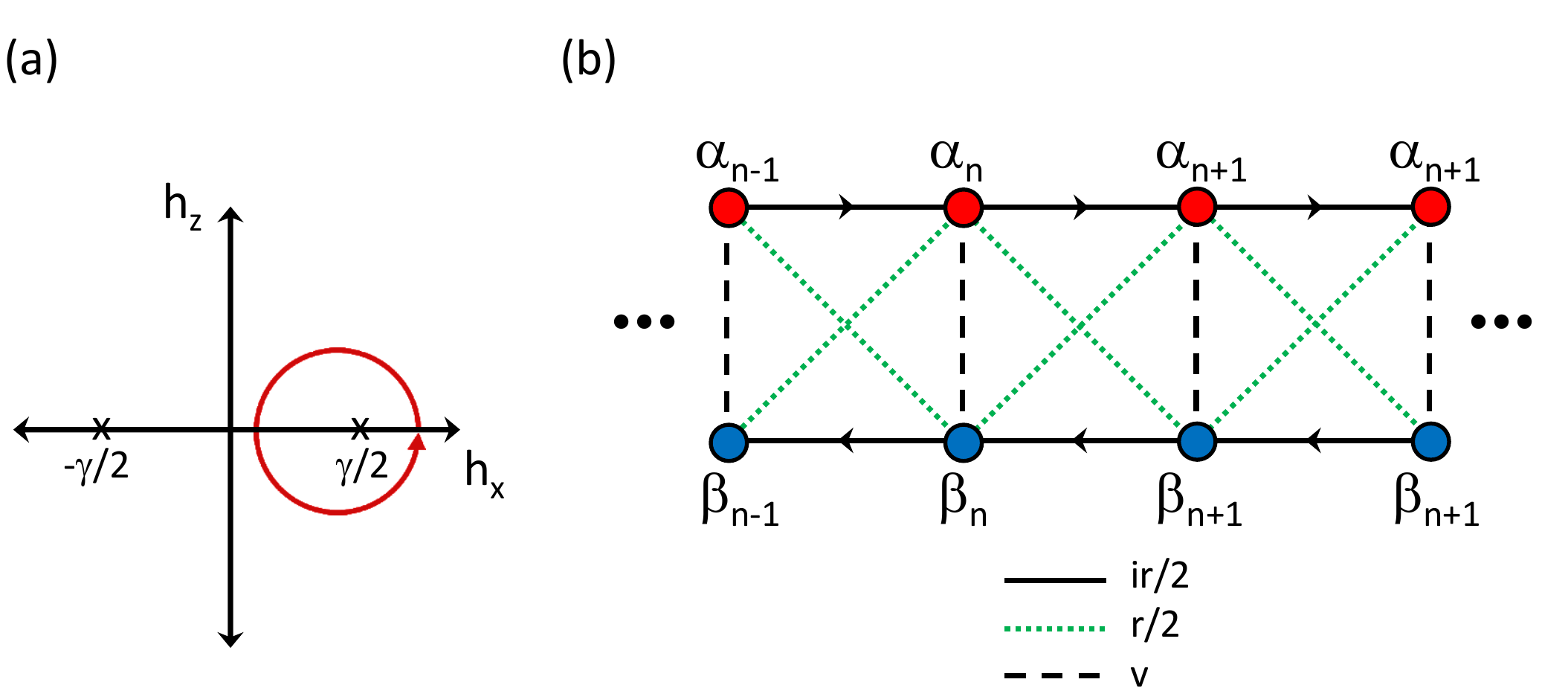}
\caption{\label{fig:model} (a) In momentum space, the Hamiltonian encircles an exceptional point at $(h_x,h_z)=(\pm\gamma/2,0)$. (b) The equivalent tight-binding model with gain on sublattice $\alpha$ and loss on sublattice $\beta$. The arrows denote the phase direction.}
\end{figure}

\emph{Model.---}
To motivate our model, we first consider a simple $2\times2$ non-Hermitian Hamiltonian,
\begin{eqnarray}
H_k&=&h_x\sigma_x + \left(h_z + \frac{i\gamma}{2}\right)\sigma_z. \label{eq:Hk}
\end{eqnarray}
where $\sigma_x,\sigma_z$ are Pauli matrices. The eigenvalues are degenerate when $(h_x,h_z)=(\pm\gamma/2,0)$. Such non-Hermitian degeneracies are called exceptional points \cite{heiss12}. We vary the parameters $h_x,h_z$ so as to encircle an exceptional point \cite{milburn15}:
\begin{eqnarray}
h_x&=&v+r\cos k, \quad\quad h_z=r\sin k,\label{eq:hz}
\end{eqnarray}
where $k$ is an external parameter for now. As $k$ increases, we make a circular trajectory in $h_x,h_z$ space [Fig.~\ref{fig:model}(a)]. When \mbox{$v-r<\pm\gamma/2<v+r$}, $H_k$ encircles the exceptional point at $(\pm\gamma/2,0)$. Without loss of generality, we assume that $r$ is positive, while $v$ can take any sign.


\begin{figure*}[t]
\centering
\includegraphics[width=6.5 in,trim=0in 4in 0.3in 4.1in,clip]{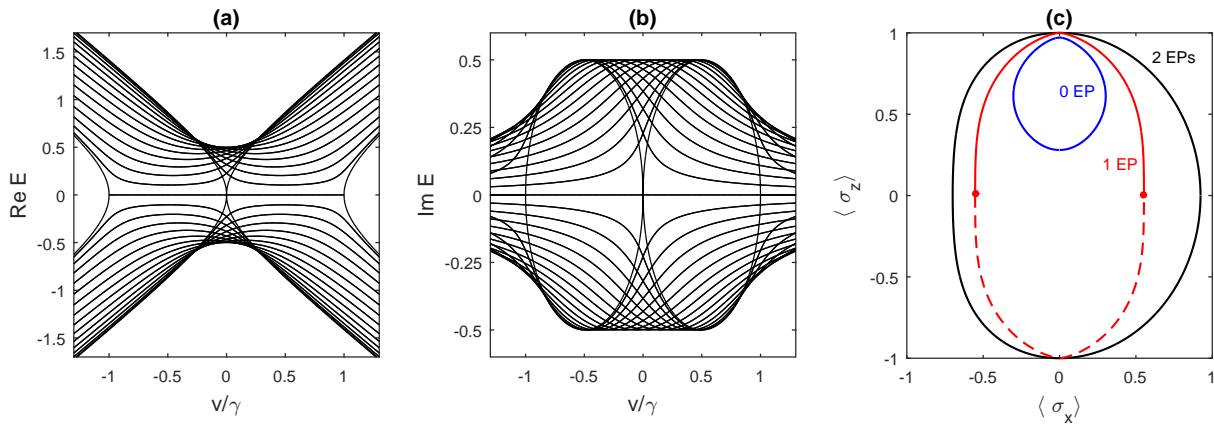}
\caption{\label{fig:periodic} Chain with periodic boundaries, $N=30$ unit cells, and $r=0.5\gamma$. (a) Real and (b) imaginary parts of the spectrum. (c) Trajectory of eigenvector $u_{k,+}$ in the Brillouin zone for $v=0.3\gamma$ when $H_k$ encircles 0 exceptional points (EPs) (blue, $r=0.18\gamma$), 1 exceptional point (red, $r=0.3\gamma$), and 2 exceptional points (black, $r=\gamma$). Solid lines denote $k\in[0,2\pi)$. Dashed line denotes $k\in[2\pi,4\pi)$.}
\end{figure*}

Now, we find the tight-binding model whose momentum-space Hamiltonian is $H_k$. The tight-binding model is given by a non-Hermitian Hamiltonian $H$. As shown in Fig.~\ref{fig:model}(b), $H$ is a one-dimensional lattice with long-range hopping as well as gain and loss. To implement $H$, the physical system we have in mind is a classical system of optical resonators with complex amplitudes $\alpha_n,\beta_n$ that obey the equations of motion
\begin{eqnarray}
\dot{\alpha}_n&=&\frac{\gamma}{2}\alpha_n-iv\beta_n+\frac{r}{2}(\alpha_{n-1}-\alpha_{n+1})-\frac{ir}{2}(\beta_{n-1}+\beta_{n+1})\nonumber\\
\dot{\beta}_n&=&-\frac{\gamma}{2}\beta_n-iv\alpha_n-\frac{r}{2}(\beta_{n-1}-\beta_{n+1})-\frac{ir}{2}(\alpha_{n-1}+\alpha_{n+1})\nonumber\\ \label{eq:H}
\end{eqnarray}
for $n=1,\ldots,N$, where $N$ is the number of unit cells, so there are a total of $2N$ sites. Equation \eqref{eq:H} is mathematically equivalent to evolving a Schr\"odinger equation with $H$ \cite{guo09,ruter10,regensburger12,zeuner15,zhen15}. $\gamma$ is the non-Hermitian gain and loss on the $\alpha$ and $\beta$ sublattices, respectively. $v,r$ denote the Hermitian hopping between sites.

Equation \eqref{eq:H} is valid when the light fields are weak, such that the dynamics are linear. For strong light fields, the gain medium saturates, and the dynamics become nonlinear \cite{dutta13}. We are interested in the linear regime, when the system is equivalent to a tight-binding model. Note that an experiment would also have noise [not shown in Eq.~\eqref{eq:H}].

$H$ is similar to Hofstader's model of a particle moving on a discrete lattice in a magnetic field, since the particle accumulates a phase of $\pi$ when it goes around a plaquette \cite{hofstadter76}. Later on, we discuss how to implement Eq.~\eqref{eq:H} experimentally in an array of coupled resonator optical waveguides \cite{hafezi11}. 

$H$ has two important symmetries. The first is chiral symmetry: letting $\Gamma=\bigoplus_n \sigma_{y,n}$, $\Gamma H\Gamma=-H$. So if $H$ has an eigenvector $u_n$ with eigenvalue $E_n$, then $\Gamma u_n$ is also an eigenvector with eigenvalue $-E_n$. There is also parity-time ($\mathcal{PT}$) symmetry: letting $\mathcal{P}=\bigoplus_n\sigma_{x,n}$ and $\mathcal{T}i\mathcal{T}^{-1}=-i$, $\mathcal{PT}H\mathcal{T}^{-1}\mathcal{P}^{-1}=H$. This means that the eigenvalues of $H$ can be real, despite the non-Hermiticity \cite{bender98}.


\emph{Periodic boundary conditions.---}
For a periodic chain, the Hamiltonian $H$ is diagonalized in momentum space as $H_k$, given in Eqs.~\eqref{eq:Hk}--\eqref{eq:hz}, where $k\in[0,2\pi)$. The eigenvalues of $H_k$ are
\begin{eqnarray}
E_k&=&\pm\sqrt{(v+r\cos k)^2 + (r\sin k+i\gamma/2)^2} \label{eq:Ek}
\end{eqnarray}
The real part of $E_k$ is gapped when $||v|-r|>\gamma/2$. The imaginary part is gapped when $||v|+r|<\gamma/2$. Figure \ref{fig:periodic} shows an example spectrum.
The eigenvectors of $H_k$ are
\begin{eqnarray}
u_{k,+} = \left(\begin{array}{c}\cos\theta/2 \\ -\sin\theta/2 \end{array}\right), \quad u_{k,-} = \left(\begin{array}{c}\sin\theta/2 \\ \cos\theta/2 \end{array}\right), \label{eq:uk}
\end{eqnarray}
where $\tan\theta=-(v+r\cos k)/(r\sin k + i\gamma/2)$.

\begin{figure*}[t]
\centering
\includegraphics[width=6.5 in,trim=0in 4in 0.3in 4.1in,clip]{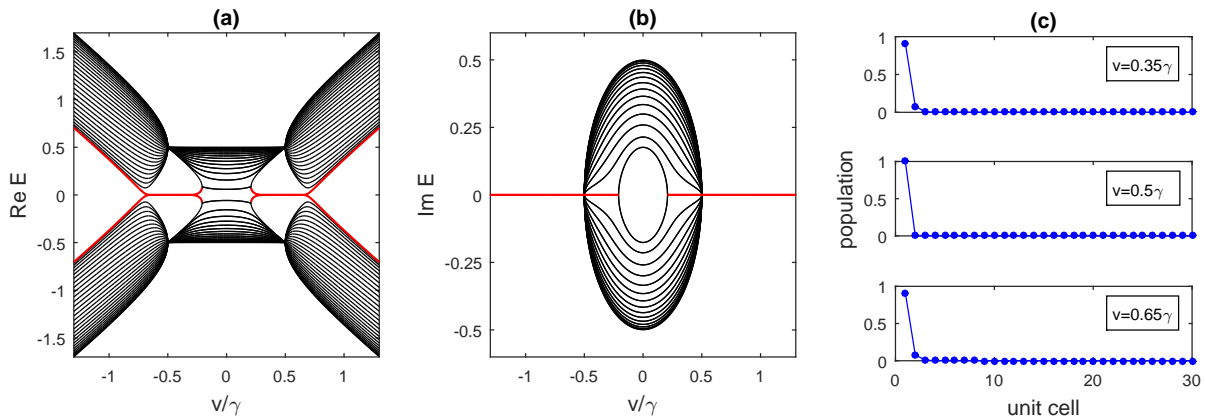}
\caption{\label{fig:open} Chain with open boundaries, $N=30$ unit cells, and $r=0.5\gamma$. (a) Real and (b) imaginary parts of the spectrum. Red lines follow the $E=0$ eigenvalue. (c) Zero-energy eigenvector for different values of $v$.}
\end{figure*}

The eigenvalues and eigenvectors of $H_k$ are $4\pi$ periodic in $k$ when $H_k$ encircles one exceptional point \cite{heiss12,uzdin11}. This is a well-known feature of non-Hermitian systems and is due to the square root in Eq.~\eqref{eq:Ek} and the half angles in Eq.~\eqref{eq:uk}. After circling once around an exceptional point, the two eigenvectors exchange values; to come back to the initial value, one must circle twice.

Now, we calculate the winding number of the eigenvectors. We follow the trajectory of $\langle\sigma_x\rangle, \langle\sigma_z\rangle$ for an eigenvector as $k$ sweeps through the Brillouin zone and see whether it winds around the origin \cite{rudner09}. As seen in Fig.~\ref{fig:periodic}(c), the winding number depends on whether $H_k$ encircles an exceptional point. When $H_k$ does not encircle an exceptional point, the winding number is 0. When $H_k$ encircles one exceptional point, the eigenvector does wind around the origin, but $k$ must sweep through $4\pi$ in order to close the trajectory; thus, the winding number has a \emph{fractional} value of 1/2 \cite{viyuela15}. When $H_k$ encircles both exceptional points, the winding number is 1. (These results can be proven analytically.)




To experimentally observe the fractional winding number, one can modulate the hopping amplitudes in time to adiabatically sweep through the Brillouin zone (see Supplemental Material for details). An alternative approach may be to use Bloch oscillations \cite{longhi09}.

Since a periodic chain has nontrivial topology, we next investigate whether there are zero-energy edge states in a chain with open boundaries. However, there is a problem: Eq.~\eqref{eq:Ek} says that when one exceptional point is encircled, $E_k$ is gapless in both real and imaginary parts. This is worrisome because it precludes the existence of zero-energy edge states, which usually require a band gap. 



\emph{Open boundary conditions.---}
Figure \ref{fig:open} shows the spectrum for an open chain. There are several remarkable features of this spectrum:
\begin{itemize}
\item A gap opens up in the spectrum's real part in the vicinity of $v=\gamma/2$, dividing most of the eigenvalues into two distinct bands [Fig.~\ref{fig:open}(a)].
\item Within the band gap, there is an $E=0$ eigenvalue, which is twofold degenerate but \emph{defective} \cite{golub96}. This eigenvalue is associated with an eigenvector and a generalized eigenvector. Under time evolution, the eigenvector dominates over the generalized eigenvector, so the latter is unimportant to the long-time dynamics.
\item The eigenvector for $E=0$ is localized either on the left edge (when $v>0$) or the right edge (when $v<0$) [see Fig.~\ref{fig:open}(c)]. The edge state is protected by chiral symmetry, and it appears when the gap opens and disappears when the gap closes.
\item For $|v|\geq\gamma/2$, the spectrum is purely real [Fig.~\ref{fig:open}(b)], i.e., $\mathcal{PT}$ symmetry is preserved, in contrast to a periodic chain  \cite{bender98}.
\end{itemize}


\emph{Open chain: case of $v=\gamma/2$.---}
We discuss, in detail, the case of $v=\gamma/2$, where $H$ can be solved exactly. We seek the eigenvalues of $H$, as well as their algebraic and geometric multiplicities \cite{golub96}. It is more convenient to deal with $H^2$, which is block upper triangular. It is easy to show that the characteristic polynomial of $H^2$ is $f(\lambda) = \lambda^2 (\lambda-r^2)^{2N-2}$, which implies that $H$ has eigenvalues $E=0,r,-r$ with algebraic multiplicities $2,N-1,N-1$, respectively. The Jordan normal form indicates that the geometric multiplicities of all three eigenvalues are 1. Thus, $H$ is highly defective at $v=\gamma/2$. Note that the eigenvalues are real.

The eigenvector for $E=0$ is the edge state
\begin{eqnarray}
u_0&=&(i,1,0,0,\ldots)^T, \label{eq:eigenvec_e0}
\end{eqnarray}
in the basis $\alpha_1,\beta_1,\alpha_2,\beta_2,\ldots$. So $u_0$ is localized on the left-most unit cell. It is its own chiral partner: $\Gamma u_0=-u_0$. Physically, this state has zero energy because of destructive interference between the hopping and non-Hermiticity.

The generalized eigenvector $u_0'$ for $E=0$ is given by $Hu_0'=u_0$ \cite{golub96}:
\begin{eqnarray}
u_0'&=&\left(\frac{2}{\gamma},0,-\frac{r}{\gamma^2},-\frac{ir}{\gamma^2},\frac{r^2}{\gamma^3},\frac{ir^2}{\gamma^3},\ldots\right)^T. \label{eq:geneigenvec_e0}
\end{eqnarray}
Since $Hu_0'=u_0$, population in $u_0'$ is transferred to $u_0$ during the time evolution. Note that $u_0'$ is also localized on the left edge.

For $v=-\gamma/2$, $H$ can be similarly solved: $u_0$ and $u_0'$ are similar to Eqs.~\eqref{eq:eigenvec_e0}--\eqref{eq:geneigenvec_e0} but localized on the right.


\emph{Open chain: case of $v\neq\gamma/2$.---}
As $v$ deviates from $\gamma/2$, the bands are no longer degenerate, and the band gap narrows and eventually closes. There is still a defective $E=0$ eigenvalue because it is protected by chiral symmetry [Fig.~\ref{fig:open}(c)]. However, when the band gap closes, the $E=0$ eigenvalue splits into two distinct eigenvalues that join the upper and lower bands [Fig.~\ref{fig:open}(a)].



Strictly speaking, for finite $N$, the $E=0$ eigenvalue is defective only when $v=\gamma/2$. However, we find numerically that for a range of $v$ around $\gamma/2$, $H$ has one vanishingly small singular value \cite{golub96}, which decreases as $N$ increases (see Supplemental Material). This indicates that for large $N$, the $E=0$ eigenvalue remains defective for a range of $v$.

In the Supplemental Material, we show that the $E=0$ eigenvalue is robust to disorder due to protection by chiral symmetry. The eigenvalue remains at $E=0$ until the disorder is strong enough to close the band gap.




\emph{Discussion.---}
The general time-dependent solution to Eq.~\eqref{eq:H} involves the eigenvectors and generalized eigenvector of $H$. Consider the regime when the $E=0$ eigenvalue exists and is defective. In this case, the coefficient of $u_0$ increases linearly in time, while the coefficient of $u_0'$ is independent of time \cite{boyce12}. Thus, as time increases, the population in the $E=0$ subspace is dominated by $u_0$.

In a typical topological insulator, there are a pair of $E=0$ eigenvectors (one on the left and one on the right), and they are equally important to the dynamics \cite{asboth16}. In our non-Hermitian model, due to the defectiveness of the $E=0$ eigenvalue, it has one eigenvector and one generalized eigenvector (both on the same edge). However, only the eigenvector is present in the long-time-scale dynamics. Thus, the generalized eigenvector is dynamically unstable, and there is only one dynamically stable edge state. We note that in the Hermitian limit ($\gamma=0$), our model behaves entirely like a typical topological insulator.


Our non-Hermitian model is highly sensitive to boundary conditions. A periodic chain has a complex spectrum and a nonzero winding number, while an open chain can have a real spectrum but has no winding number. An open chain also has a band gap not present in a periodic chain. These differences are because an open chain is much more defective than a periodic chain.

We note that the Su-Schrieffer-Heeger model also has a single zero-energy state when the number of sites is odd \cite{sirker14}. Because of the incompleteness of a unit cell, this state exists even when the gap closes. This state is not defective because the algebraic and geometric multiplicities are both 1. In contrast, our model's zero-energy state is defective and disappears when the gap closes. Our zero-energy state originates from the non-Hermiticity instead of the peculiarity of an incomplete unit cell.



\emph{Experimental implementation.---}
Equation \eqref{eq:H} can be implemented with an array of coupled resonator optical waveguides similar to Ref.~\cite{hafezi11}. In this setup, each site is a resonator, and waveguides between resonators allow photons to hop between sites. One would design an array of resonators with waveguide connections as in Fig.~\ref{fig:model}(b). The phase of a hopping amplitude can be tuned by making the corresponding waveguide asymmetric; in this way, one can engineer the imaginary hopping amplitudes in $H$. To avoid cross talk between the diagonal waveguides that cross each other, one diagonal should be below the other.


\begin{figure}[t]
\centering
\includegraphics[width=3.3 in,trim=1.6in 3.4in 1.7in 3.4in,clip]{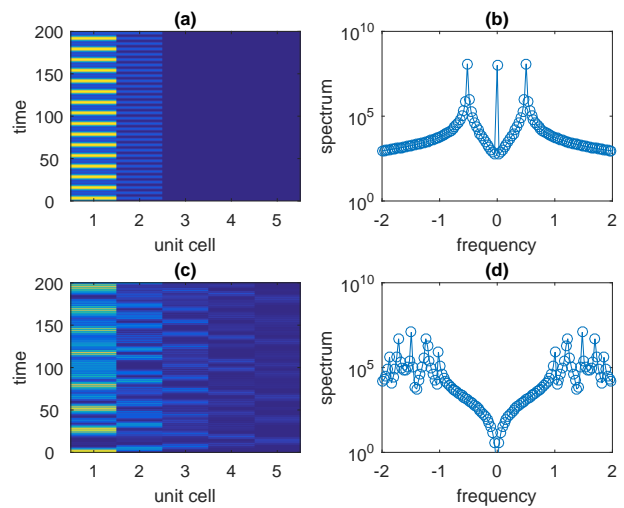}
\caption{\label{fig:fourier}Open chain of $N=5$ unit cells, starting with excitation on $\alpha_1$, (a,b) when zero-energy state is present ($v=0.5\gamma,r=0.5\gamma$) and (c,d) when it is absent ($v=1.5\gamma,r=0.5\gamma$). (a,c) Population on each unit cell over time, plotted using color scale. (b,d) Fourier spectrum of $\alpha_1$.}
\end{figure}

In principle, the gain on the $\alpha$ sublattice can be obtained by incorporating a pumped gain medium as in Ref.~\cite{ruter10}. In practice, it is easier to let both sublattices be lossy but with more loss on the $\beta$ sublattice \cite{zeuner15,guo09}; such a passive setup avoids the complication of the gain medium. The physics is still described by Eq.~\eqref{eq:H} but on top of a background of decay. 

When $|v|\geq\gamma/2$, $H$ has real eigenvalues, so the evolution is oscillatory (Fig.~\ref{fig:fourier}). It is easy to detect the $E=0$ state by exciting the left edge and measuring the frequency spectrum of the subsequent evolution. When the $E=0$ state is present, the spectrum has a peak at zero frequency [Fig.~\ref{fig:fourier}(b)]. When the $E=0$ state is absent, there is no peak at zero frequency [Fig.~\ref{fig:fourier}(d)].



\emph{Conclusion.---}
We have shown that non-Hermiticity breaks the usual bulk-boundary scenario. The new features are due to the fact that a non-Hermitian matrix can be defective, while a Hermitian matrix is always diagonalizable. In the future, it would be interesting to consider transport properties in the presence of a potential gradient \cite{longhi09} or scattering properties, especially when the Brillouin zone is $4\pi$ periodic. One should also extend the model to two dimensions to see whether the Chern number can be fractional and whether there is still only one dynamically stable edge state.

We thank P. Rabl and Y.N. Joglekar for useful discussions.

\bibliography{edge}

\clearpage
\appendix

\begin{center}
\textbf{Supplemental Material}
\end{center}
\section{Robustness to disorder}

We now show that the zero-energy edge state is robust to disorder that preserves chiral symmetry. The $E=0$ eigenvalue remains in the spectrum until the disorder is large enough that the band gap closes. We allow $r,v,\gamma$ to be different for each unit cell:
\begin{eqnarray}
H&=&\sum_n \Big[\frac{ir_n}{2}( a^\dagger_{n+1} a_n -  a^\dagger_n a_{n+1} ) - \frac{ir_n}{2}( b^\dagger_{n+1} b_n - b^\dagger_n b_{n+1} ) \nonumber\\
&& + \frac{r_n}{2}(b^\dagger_n a_{n+1} + a_{n+1}^\dagger b_n) + \frac{r_n}{2}(b^\dagger_{n+1} a_n + a^\dagger_n b_{n+1}) \nonumber\\
&& + v_n(b^\dagger_n a_n + a^\dagger_n b_n) + \frac{i\gamma_n}{2}(a^\dagger_n a_n - b^\dagger_n b_n)\Big]. \label{eq:H_disorder}
\end{eqnarray}
$H$ still has chiral symmetry: $\Gamma H\Gamma=-H$ where $\Gamma=\bigoplus_n \sigma_{y,n}$. We note that the $r_n$ in the first line of Eq.~\eqref{eq:H_disorder} can be different from the $r_n$ in the second line and still preserve chiral symmetry.

\begin{figure}[t]
\centering
\begin{tabular}{c}
\includegraphics[width=3.3 in,trim=1.5in 3.9in 1.8in 4in,clip]{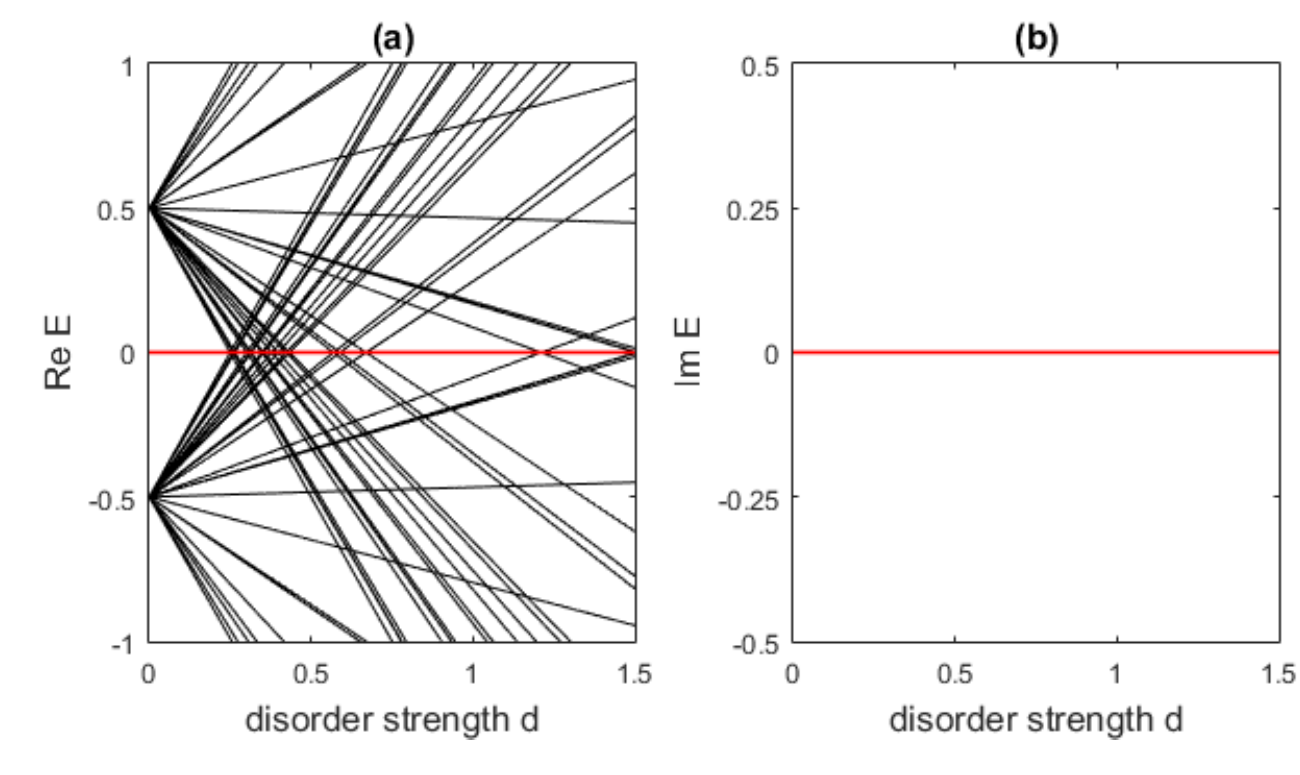}\\
\includegraphics[width=3.3 in,trim=1.5in 3.9in 1.8in 4in,clip]{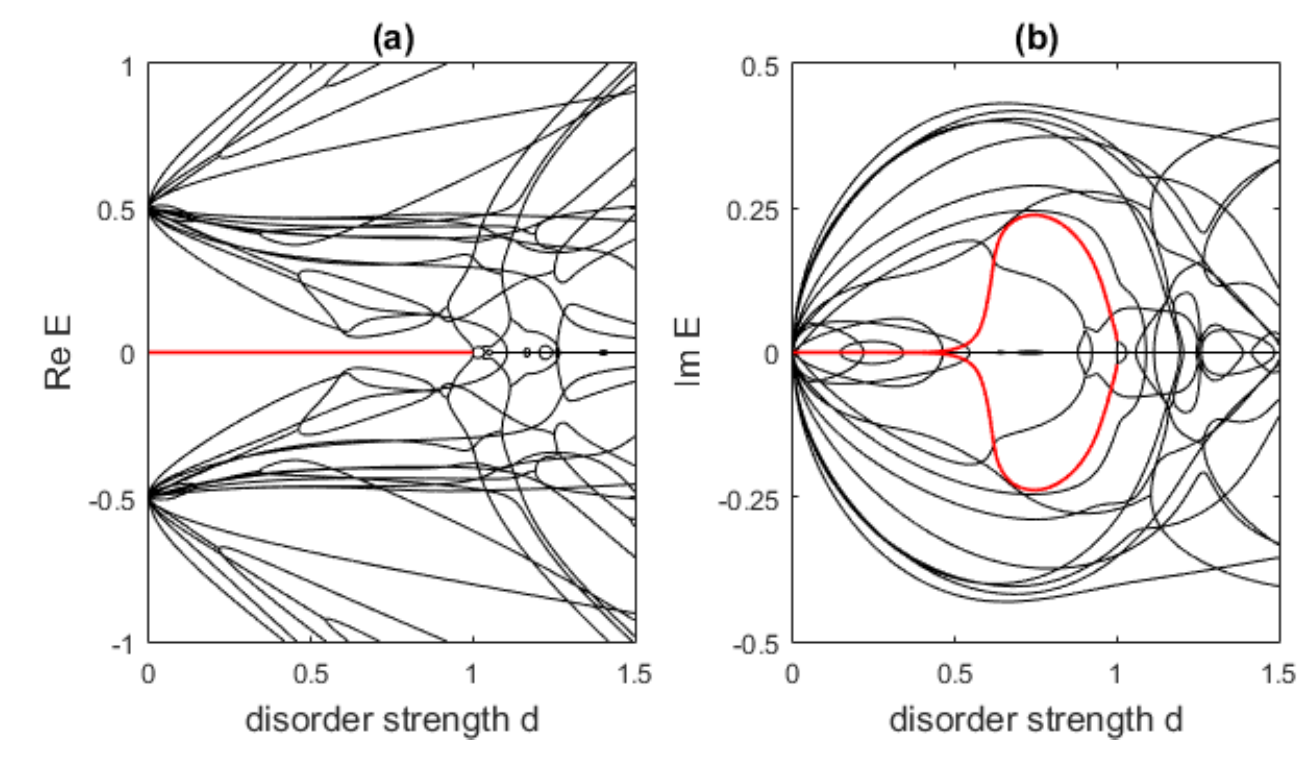}\\
\includegraphics[width=3.3 in,trim=1.5in 3.9in 1.8in 4in,clip]{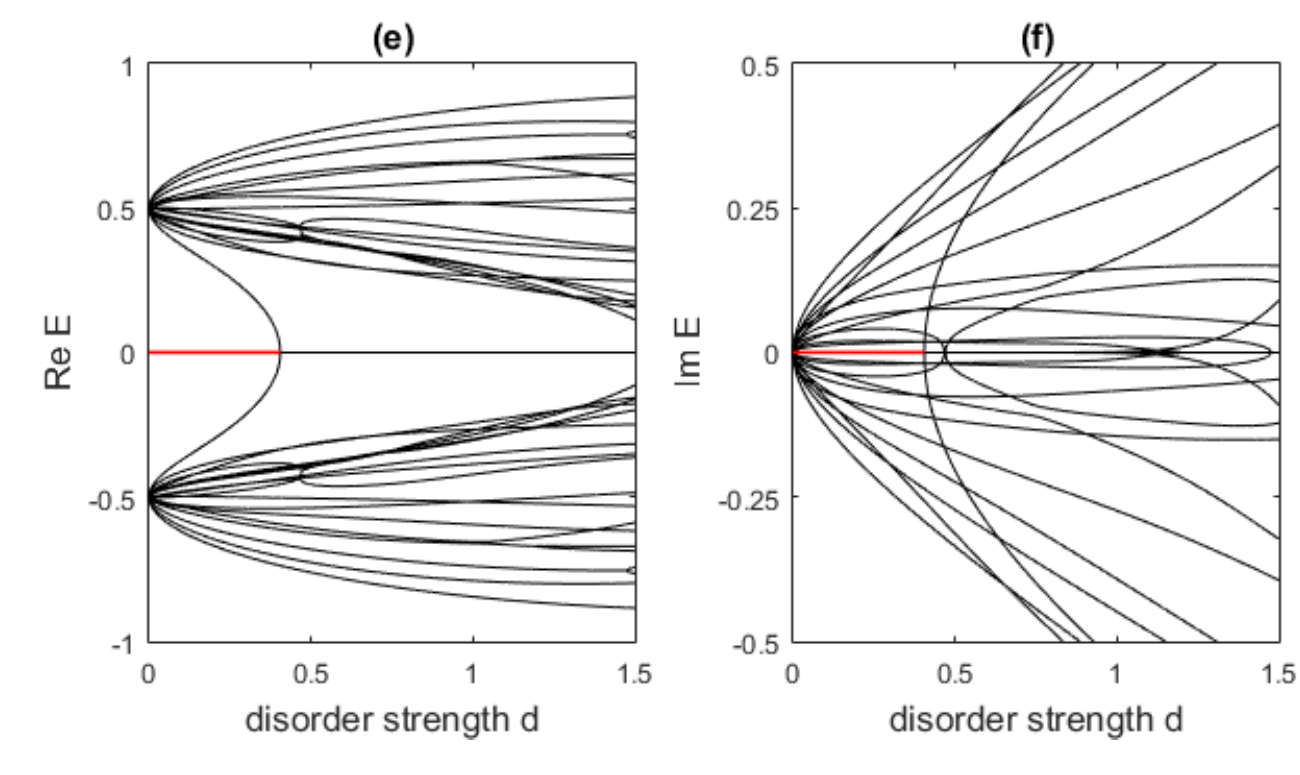}
\end{tabular}
\caption{\label{fig:disorder}Open chain with (a,b) disorder in $r$, (c,d) disorder in $v$, and (e,f) disorder in $\gamma$. (a,c,e) Real and (b,d,f) imaginary parts of the spectrum. Red lines follow the $E=0$ eigenvalue. Parameters are $N=30$, $r=0.5$, $v=0.5$, $\gamma=1$.}
\end{figure}

Figure \ref{fig:disorder}(a,b) shows an example spectrum with disorder in $r$: $r_n = r + d\,\epsilon_n$, where $\epsilon_n$ is a uniform random variable on $[-1,1]$, and $d$ is the disorder strength. The $E=0$ eigenvalue is completely robust to disorder in this case, even when the band gap closes; its eigenvector remains localized on the left-most unit cell. The bands are affected by disorder, although all eigenvalues stay real.

Figure \ref{fig:disorder}(c,d) shows an example spectrum with disorder in $v$: $v_n = v + d\,\epsilon_n$. The $E=0$ eigenvalue is robust to disorder until the real gap narrows at $d\approx0.5$, at which point it splits into two imaginary eigenvalues. Those eigenvalues undergo exceptional points with other eigenvalues at $d\approx1$.

Figure \ref{fig:disorder}(e,f) shows an example spectrum with disorder in $\gamma$: $\gamma_n = \gamma + d\,\epsilon_n$. The $E=0$ eigenvalue is robust to disorder until the real gap closes at $d\approx0.4$, at which point the eigenvalue has an exceptional point with other eigenvalues.

Thus, the $E=0$ eigenvalue is robust to disorder as long as chiral symmetry is preserved and the gap does not close. (If the disorder did not preserve chiral symmetry, the eigenvalue would immediately deviate from $E=0$ for small $d$.)

\section{Experimental detection of fractional winding number}

The most direct way to see the $4\pi$ periodicity of the eigenvectors is to modulate the hopping amplitudes in time. We multiply the hopping amplitudes by a phase factor $\phi(t)$:
\begin{eqnarray}
H&=&\sum_n \Big[\frac{ir}{2}(e^{-i\phi} a^\dagger_{n+1} a_n - e^{i\phi} a^\dagger_n a_{n+1} ) \nonumber\\
&& - \frac{ir}{2}(e^{-i\phi} b^\dagger_{n+1} b_n - e^{i\phi} b^\dagger_n b_{n+1} ) \nonumber\\
&& + \frac{r}{2}(e^{-i\phi} b^\dagger_{n+1} a_n + e^{i\phi} a^\dagger_n b_{n+1}) \nonumber\\
&& + \frac{r}{2}(e^{-i\phi} a_{n+1}^\dagger b_n + e^{i\phi} b^\dagger_n a_{n+1}) \nonumber\\
&& + v(b^\dagger_n a_n + a^\dagger_n b_n) + \frac{i\gamma}{2}(a^\dagger_n a_n - b^\dagger_n b_n)\Big]. \label{eq_app:H}
\end{eqnarray}
This phase modulation can be implemented in an optical experiment like Ref.~\cite{hafezi11} by adding a Kerr medium to one branch of the waveguides, allowing one to tune the index of refraction using an external source of light. In this way, the phase of a hopping amplitude can be changed \emph{in situ}.

In momentum space, Eq.~\eqref{eq_app:H} becomes:
\begin{eqnarray}
H_k&=&h_x\sigma_x + \left(h_z + \frac{i\gamma}{2}\right)\sigma_z. \label{eq_app:Hk} \\
h_x&=&v+r\cos (k+\phi), \quad\quad h_z=r\sin (k+\phi),\label{eq_app:hz}
\end{eqnarray}
For a given $k$, increasing $\phi$ effectively sweeps through the Brillouin zone. By changing $\phi$ adiabatically, one can directly observe the $4\pi$ periodicity of the eigenvectors. Equations \eqref{eq_app:Hk} and \eqref{eq_app:hz} have been studied in Refs.~\cite{milburn15,uzdin11}.

\begin{figure}[t]
\centering
\includegraphics[width=3.4 in,trim=1.6in 3.75in 1.6in 4in,clip]{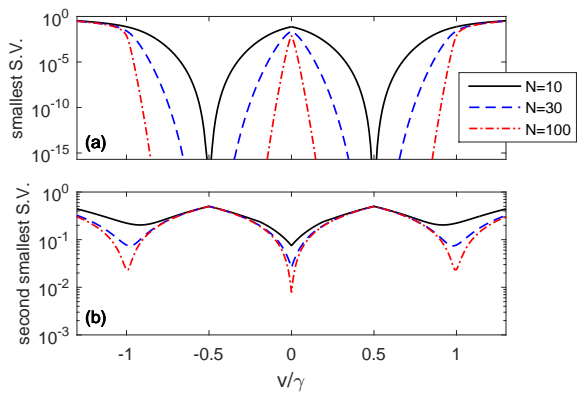}
\caption{\label{fig:singular}Singular value decomposition for an open chain with $r=0.5\gamma$ and various $N$. (a) Smallest and (b) second smallest singular values. Numerical precision here is $2\times 10^{-16}$.}
\end{figure}

For example, suppose the system is initialized in the $u_{k,-}$ eigenstate with $k=0$. Suppose also that $v-r<\gamma/2<v+r$, so one exceptional point is encircled. Starting from $\phi=0$, we increase $\phi$ in time. When $\phi=2\pi$, the system does \emph{not} return to the original eigenstate, but it will be in $u_{k,+}$. In contrast, if the exceptional point is not encircled, then the system will return to the original eigenstate. In this way, one can clearly see the $4\pi$ periodicity due to the exceptional point.

We note that the switching of the eigenvectors can depend on whether the trajectory around the exceptional point is clockwise or counter-clockwise \cite{milburn15,uzdin11}.

\section{Range of defectiveness}

Here, we provide numerical evidence that Eq.~(3) of the main text has a defective $E=0$ eigenvalue for a range of $v$. Figure \ref{fig:singular} shows the smallest and second smallest singular values of $H$. For a range of $v$ around $\pm\gamma/2$, there is one vanishingly small singular value, and it decreases as $N$ increases. Thus, for large $N$, the geometric multiplicity of the $E=0$ eigenvalue is 1 for a range of $v$. Since its algebraic multiplicity is 2, the $E=0$ eigenvalue is defective in this range.

\end{document}